# Resistance Drift in Melt-Quenched Ge$_2$Sb$_2$Te$_5$ Phase Change Memory Line Cells at Cryogenic Temperatures

A B M Hasan Talukder[z], Md Tashfiq Bin Kashem, Raihan Khan, Faruk Dirisaglik, Ali Gokirmak, and Helena Silva

Department of Electrical and Computer Engineering, University of Connecticut, Storrs, CT 06269, USA

[z]E-mail: talukder@uconn.edu

**Abstract**

We characterized resistance drift in phase change memory devices in the 80 K to 300 K temperature range by performing measurements on 20 nm thick, ~70-100 nm wide lateral Ge$_2$Sb$_2$Te$_5$ (GST) line cells. The cells were amorphized using 1.5 – 2.5 V pulses of ~50 - 100 ns duration leading to ~0.4 - 1.1 mA peak reset currents resulting in amorphized lengths between ~50 and 700 nm. Resistance drift coefficients in the amorphized cells are calculated using constant voltage measurements starting as fast as within a second after amorphization and for 1 hour duration. The drift coefficients vary between ~0.02 and 0.1 with significant device-to-device variability and variations during the measurement period. At lower temperatures (higher resistance states) some devices show a complex dynamic behavior, with the resistance repeatedly increasing and decreasing significantly over periods in the order of seconds.

**Introduction**

Electronic phase change memory (PCM) can potentially bridge the gap in density and speed between dynamic random-access memory (DRAM) and flash storage [1–4]. PCM offers fast read/write times (~10 – 100 ns), high endurance (~10$^{12}$), and long data retention (10 years at 210°C) [5–7]. This high-density 1S1R (1 switch, 1 resistor) technology is based on a chalcogenide material that forms the active region of the cell and switches between the highly conductive crystalline phase and highly resistive amorphous phase with two to four orders of magnitude resistivity contrast [8,9] which is attributed to the structural order, bonding length and angle and carrier concentration difference between the phases [4,10,11]. Ge$_2$Sb$_2$Te$_5$ (GST) has been the most studied alloy [12] for PCM because of its fast crystallization speed, relatively lower melting point (~858K [13]), high resistivity contrast between amorphous and crystalline phases [14], high thermodynamic stability [15], and high endurance (up to ~10$^{12}$ cycles) [7,16]. Even though the cell operation is rather complicated compared to conventional electronic devices that remain close to room temperature, and the behavior of the materials is not fully understood yet, PCM is proven to be a reliable non-volatile memory technology that can be integrated with CMOS at the back-end-of-the line [17]. However, resistance drift in the amorphous state, generally attributed to structural relaxation of the material [18–25], remains a significant challenge for implementation of multi-level cell (MLC) operation. Once amorphized, the resistance of the material increases over time approximately following a power law behavior [26–30]:

$$R = R_0 \left(\frac{t}{t_0}\right)^\nu \quad (1)$$

where $R$ and $R_0$ are the cell resistances at time $t$ and $t_0$ and $\nu$ is the drift coefficient.

In this work, we perform detailed I-V characterization and resistance drift monitoring of melt-quenched amorphous cells of similar dimensions. The amorphized length is estimated based on the post-pulse device resistance and resistivity of metastable melt-quenched amorphous GST.

**Device Structure and Experimental Setup**

We performed our measurements on two terminal GST line cells (Fig. 1a) with 250 nm thick bottom metal contacts (W with TiN liner) on 600 nm thermally grown $SiO_2$. A 20 nm thick layer of GST was deposited over the metal contacts by co-sputtering from elemental targets at room temperature resulting in as-deposited amorphous phase. The line cells are patterned using photolithography and reactive ion etching (RIE), and are capped with 15 nm

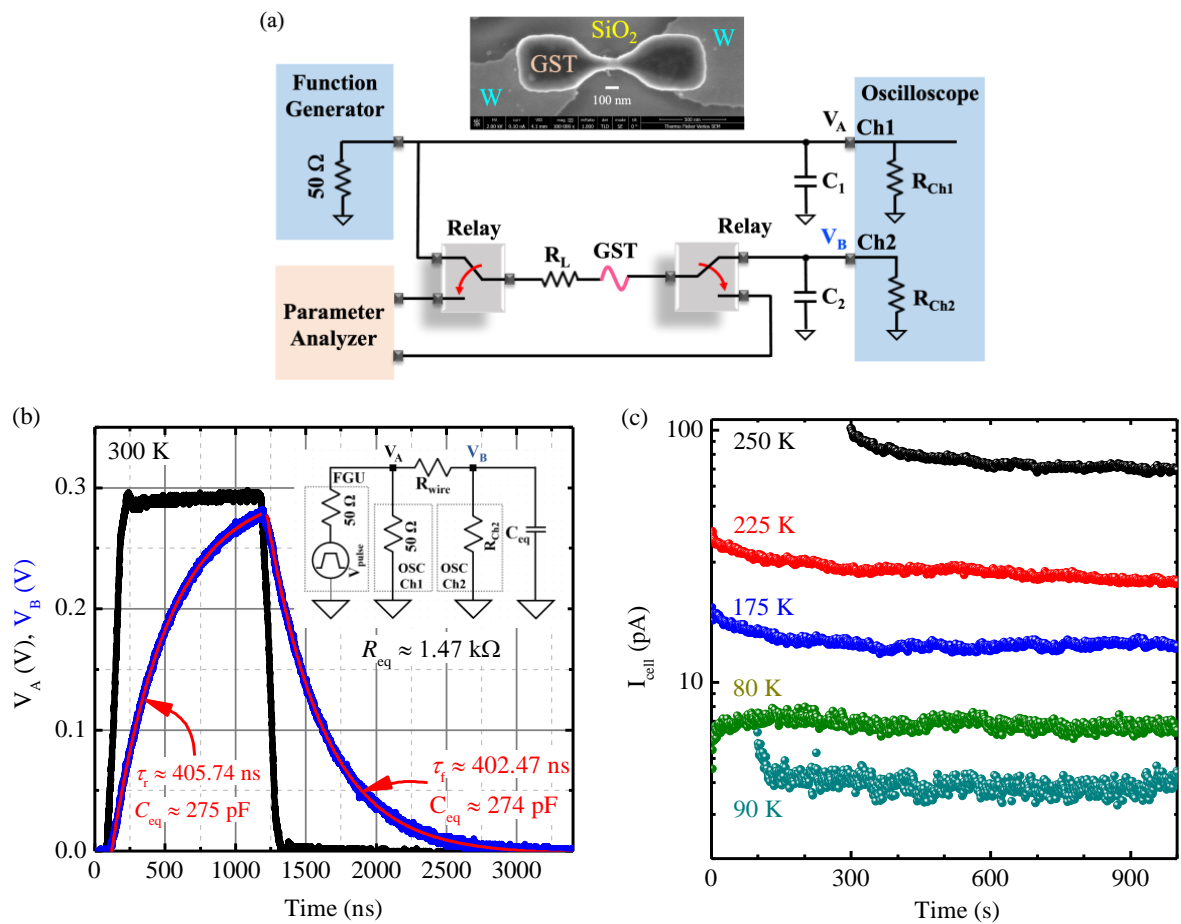

**Figure 1.** (a) Scanning electron microscopy (SEM) image of an example GST line cell with Tungsten (W) bottom contacts. The equivalent circuit schematic of the measurement setup consisting of a 2-channel relay between a function generator-oscilloscope connection and a semiconductor parameter analyser to carry on the measurements using a cryogenic probe station (Janis ST-500-UHT). The oscilloscope channel 1, with 50 Ω termination ($R_{Ch1}$ = 50 Ω), captures the applied pulse while channel 2 is used to capture the resulting current pulse from $V_B$. (b) Characterization of the setup parasitic equivalent capacitance, $C_{eq}$, with $R_{Ch2}$ = 1 MΩ with a pulse of sufficiently long duration to let $C_{eq}$ charge and discharge completely. $R_{Ch2}$ is replaced with a 50 Ω termination for device characterization. (c) The use of the 2-channel relay (Agilent 16440A) enables low leakage measurements within a second from amorphization.

SiO$_2$ deposited by plasma enhanced chemical vapor deposition (PECVD) to prevent any oxidation and/or evaporation during operation. Details of the device fabrication are available in Refs. [31,32]. Two sets of line cells are used in this work: a set of narrower devices with design dimension of W × L ≈ 72 - 86 nm × 246 - 352 nm and a set of wider devices with design dimension of W × L ≈ 100 nm × 320 - 340 nm. The cells are annealed at 675 K for ~20 minutes to crystallize GST to their hexagonal close pack (hcp) phases.

The electrical characterization setup (Fig. 1a) is controlled through a computer (LabVIEW interface) and includes a 2-channel relay (Agilent 16440A) that automatically connects the PCM cell to a parameter analyzer (Agilent 4156C) or an arbitrary waveform generator (Tektronix AFG 3102) and a digital oscilloscope (Tektronix DPO 4104). The parameter analyzer is used for I-V characterization and to monitor the resistance drift after amorphization, while the arbitrary waveform generator and the oscilloscope are used to apply the amorphization pulses and to perform the measurements during pulsing. The 2-channel relay is controlled through an Arduino I/O interface and enables fast switching (< 1 s) and low- leakage (< 10 fA) measurements. With this setup, it is possible to monitor the behavior of the device amorphized to very high resistance levels, within a second of amorphization, while in our earlier experiments we could not be perform high sensitivity measurements without manually disconnecting the coaxial cables which delayed the measurements by ~100 s [33].

To determine the limitations of the setup, we characterized the parasitic capacitance in the setup due to coaxial and triaxial cabling, connectors, and switches. We terminated the second channel of the oscilloscope at 1 MΩ and measure the time constants associated with the equivalent capacitance ($C_{eq}$) in the system from the pulse characteristics (Fig. 1b). With the time constant values extracted from the exponential fits to the rising edge and falling edge of the pulse $V_B$, we calculated the equivalent capacitance to be $C_{eq} \approx 275 \pm 1$ pF. Hence, the system rise/fall-time with 50 Ω termination is ~14 ns and the capacitive current during the amorphization pulses is found to be negligible compare to the current calculated at the 50 Ω termination. The current through the cell during reset operation is therefore calculated from the voltage across the channel 2 termination of 50 Ω ($I_{cell} \approx V_B/50$ Ω).

After characterizing the measurement setup, we measure the I-V characteristics of GST line cells prior to amorphization, after being annealed at 675 K for ~20 minutes, with a low-voltage dc sweep, typically from -0.3 to 0.3 V, to confirm the low-resistance state (expected *hcp* phase). An amorphization pulse of suitable duration and amplitude (depending on device dimensions) is then applied and the resulting waveforms are recorded. In some devices amorphization is confirmed with a subsequent low-voltage dc sweep, followed by higher voltage sweeps for transport characterization. In others, current through the device for a constant read voltage continues to be recorded to monitor the resistance drift for one hour.

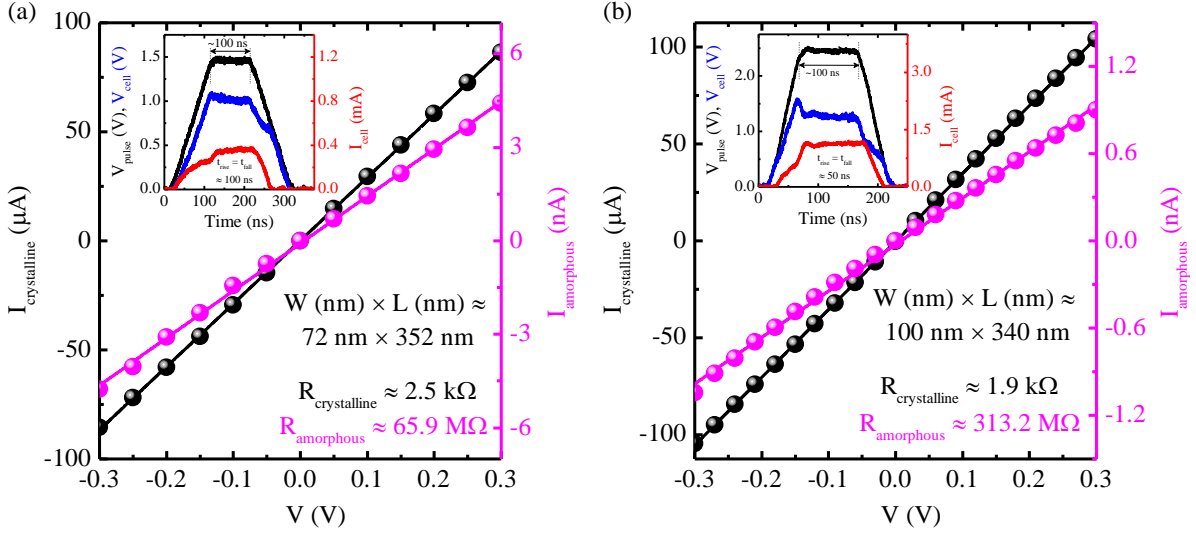

**Figure 2.** Current in the crystalline and amorphous phases measured from the low-voltage I-V characteristics before and after applying the amorphization pulse shown for an example narrow device with design width of 72 nm and design length of 352 nm (a), and for an example wider device with design W × L ≈ 100 nm × 340 nm (b). For the narrow devices, ~1.5 V amorphization pulse results in ~0.4 mA peak current through the device and resistance contrast ~$10^4$, whereas for the wider devices a larger amplitude (~2.5 V) pulse results in ~1 mA peak current and resistance contrast of ~$10^5$.

**Results and Discussion**

The initial resistance level in the crystalline phase of the GST cells under test, extracted from the I-V characteristics measured at room temperature, ranges from ~1.2 kΩ to 8.8 kΩ. With the application of a suitable amorphization pulse, the resistance increases to ~28 – 400 MΩ. Figure 2 shows how the current in amorphous phase ($I_{amorphous}$) compares with the current in crystalline phase ($I_{crystalline}$), showing amorphous-to-crystalline resistance contrast of ~$10^4 - 10^6$. The choice of amorphization pulse amplitude and duration depends on the cell dimension. For the set of narrower GST cells, we apply an amorphization pulse of ~1.5 V amplitude and ~100 ns pulse with ~100 ns rise time ($t_{rise}$) and ~100 ns fall time ($t_{fall}$) (Fig. 2a inset) whereas for the set of wider cells, the amorphization pulse is typically ~2.5 V and 100 ns with $t_{rise} = t_{fall} \approx 50$ ns (Fig. 2b inset). The narrow device starts melting at a current of ~0.3 mA at ~1.05 V while the wider device starts melting at a higher current of ~0.7 mA at ~1.55 V. The current averaged for the flat duration of the resulting current pulses is in the range of ~0.5 mA for the narrow devices and ~1 mA for the wider devices. Higher amplitude pulses typically result in breaking of the narrower cells. On the other hand, lower amplitude pulses typically fail to amorphize the wider cells.

To extract the amorphized length ($L_{amorphized}$) we use our previously measured metastable room-temperature resistivity value of ~100 Ω.cm [32] and the cell amorphous resistance level, together with the design width of the devices (confirmed to be very close to the physical width by SEM). The amorphized length increases approximately linearly with $I_{cell,peak}$ during the melt-quench and for higher currents it approaches the length of the narrow section of the GST lines

(L, as shown in Fig. 3b inset). For even higher currents, the amorphized length extends into the wider GST pads and approaches the metal-to-metal distance ($L_m$).

**Table I**. Amorphized length ($L_{amorphized}$) and low-field to high-field transition voltage and field.

| W (nm) | L (nm) | $I_{cell,peak}$ (mA) | $R_{amorphous}$ (MΩ) | $L_{amorphized}$ (nm) | $V_{tr}$ (V) | $E_{tr}$ (MV/m) |
|---|---|---|---|---|---|---|
| 72  | 352 | 0.36 ± 0.01 | 65.9 ± 0.6   | 95 ± 1  | 2.0  | 21.1 ± 0.2  |
| 76  | 246 | 0.42 ± 0.01 | 145.4 ± 0.7  | 221 ± 1 | 2.3  | 10.4 ± 0.05 |
| 76  | 256 | 0.45 ± 0.01 | 399.9 ± 2.3  | 608 ± 3 | 3.8  | 6.3 ± 0.04  |
| 86  | 246 | 0.57 ± 0.01 | 73.3 ± 0.4   | 126 ± 1 | 2.3  | 18.2 ± 0.09 |
| 86  | 256 | 0.50 ± 0.01 | 28.3 ± 1.3   | 49 ± 2  | *    | *           |
| 86  | 276 | 0.62 ± 0.01 | 120.5 ± 1.4  | 207 ± 2 | 4.2  | 20.3 ± 0.23 |
| 100 | 320 | 1.14 ± 0.02 | 301.9 ± 3.2  | 604 ± 6 | 14.8 | 24.5 ± 0.26 |
| 100 | 330 | 1.17 ± 0.02 | 353.5 ± 2.8  | 707 ± 6 | 13.7 | 19.4 ± 0.15 |
| 100 | 340 | 1.13 ± 0.02 | 313.2 ± 2.8  | 626 ± 6 | 14.0 | 22.3 ± 0.20 |

$V_{tr}$ was not observed clearly for the device (*).

The estimated amorphized length is important to characterize and better understand resistance drift and any dependences on device size. High-field I-V sweeps on amorphized cells show different transport regimes as observed before [34], having low- and high-field regimes (Fig. 3a), attributed to different dominant conduction mechanisms [35]. We observe two distinct exponential responses in the I-V characteristics: low-field response and high-field response. The low-field response can be modeled by 2D thermally activated hopping transport model described in Ref [34] and the high-field response can be modeled as

$$I_{HF} = I_{0,H}\left(e^{\alpha(V-V_H)}-1\right), \qquad V \geq V_H \tag{2}$$

where $I_{0,H}$ is the pre-factor for high-field response, α is coefficient that modulates the slope of the response, and $V_H$ is the threshold voltage at which the high-field process starts to initiate. Below $V_H$, the high-field current $I_{HF}$ approaches zero.

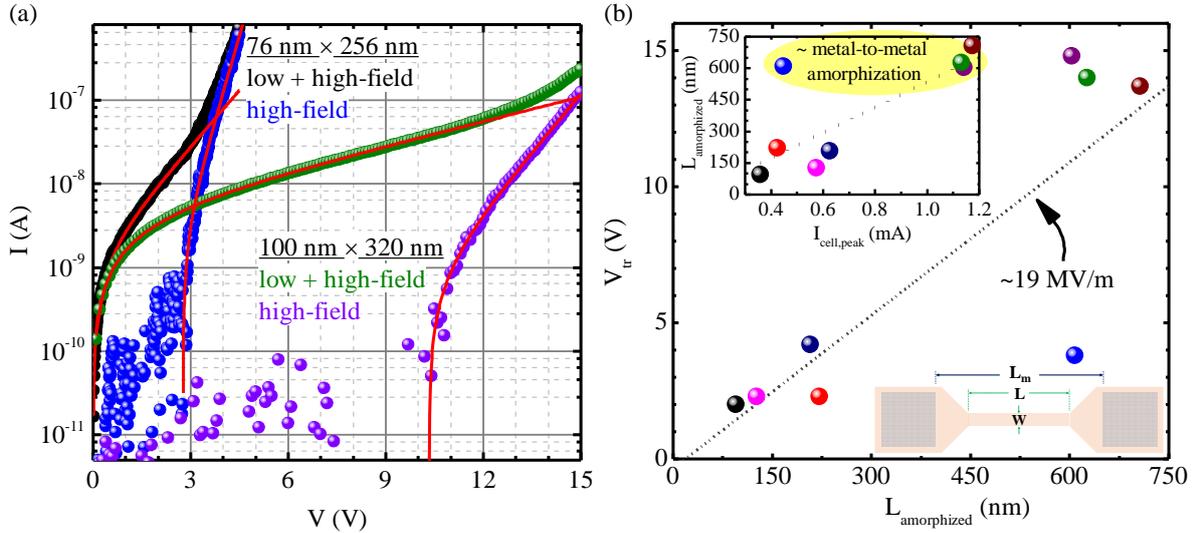

**Figure 3.** (a) I-V characteristics from the voltage sweeps high enough to result in a transition from a low-field response to a high-field response. For the set of narrow devices, the transition happens at lower voltages (also shown in the inset) at $V_{tr} \leq 4.2$ V compared to a significantly higher transition voltage ($V_{tr} \geq 12.5$ V) for the wider devices. (b) The transition voltages show an approximately linear relationship to the amorphized length of the GST cells. The linear fit to transition voltages vs the amorphized lengths gives a slope of ~19 MV/m as the transition field required for switching from low- to high-field responses. A higher $I_{cell,peak}$ during melt-quenching results in higher amorphized length reaching up to metal-to-metal length as shown in the top inset.

The observed low-to-high field transition voltages obtained from the intersecting point of the low- and high-field fits (Fig. 3a), when plotted against the amorphized lengths (Fig. 3b), show an approximate transition field of 19 MV/m, in line with previous reports of steeper field response at fields > 10 MV/m [35].

The dependence of resistance drift behavior on temperature was studied by drift measurements between 80 and 300 K, under low read field, for one hour, in 38 GST devices with W × L ≈ 66 – 88 nm × 220 – 388 nm (Fig. 4a). The drift coefficients are determined from linear fits of log($R/R_0$) vs log($t/t_0$) plots.

The resistance drift however is not monotonous, and we observe significant variations over time, during the one-hour acquisition, including periods of decreasing resistance. To illustrate the significant variation in the short-term resistance drift observed in these measurements, point-by-point resistance drift coefficients are obtained using a time-moving window of 100 and 500 seconds (Fig. 4b).

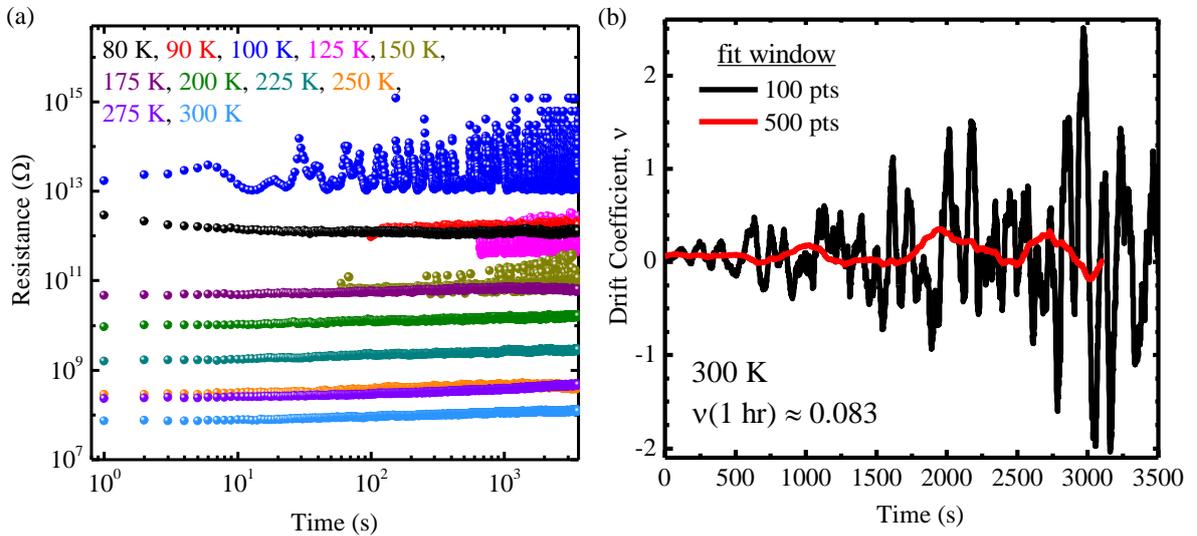

**Figure 4.** (a) Resistance of example GST cells monitored under low field (100 mV) after amorphization by applying a suitable read voltage (depending on device dimensions) for one hour, at different temperatures (amorphization and drift monitoring at the given temperatures). Since the amorphous resistance is significantly higher at lower temperatures, higher read voltages are required (10-12 V for 80 K, 3-7.5 V for 90-125 K, 0.5-1 V for 150-175 K). (b) The drift coefficient is obtained from linear fits of log(*R*) vs log(*t*) and varies depending on the fitting window chosen. Moving fit window of 100 points and 500 points out of 3600 points (1 hour drift measurements) show this variation for a cell amorphized and monitored at 300 K.

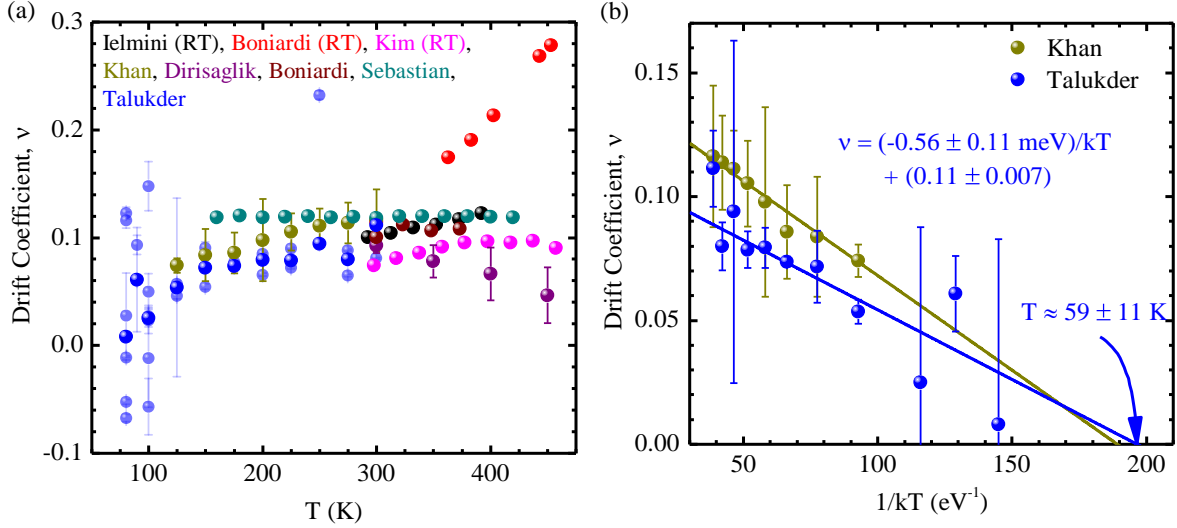

**Figure 5**. (a) Resistance drift coefficient of melt-quenched amorphous phase from this work and others [21,33,36–39]. Some of the reported drift measurements were performed at the annealing temperature ($T_R = T_A$) while others at room temperature ($T_R = $ RT). The results from this work are from 38 devices (shown in light blue spheres ○) with W × L ≈ 66 – 88 nm × 220 – 388 nm, obtained from linear fits of $\log(R/R_0)$ vs $\log(t/t_0)$ for the first 600 points after amorphization and show relatively smaller drift coefficients compared to our previously reported values from 48 devices with width (W) × length (L) ≈ 122 – 142 nm × 360 – 500 nm obtained from linear fits of $\log(R/R_0)$ vs $\log(t/t_0)$ for the first 10,000 s after amorphization [33]. (b) Linear fit of $\log(\nu)$ vs $1/kT$ shows an approximate slope of ~0.56 meV determined from the first 600 points for this work, compared to ~0.76 meV for our previous work (Ref [33]) showing a zero-drift temperature of ~59 K. The error bars correspond to the standard deviation calculated from the measurements on different devices at each temperature.

For comparison purposes, the resistance drift coefficient versus temperature in Fig. 5 are obtained from linear fits of $\log(R/R_0)$ vs $\log(t/t_0)$ for the first 600 points (first 600 s after amorphization for most of the devices) and are in general agreement to our previous longer term drift measurements (shown in Fig. 5a, together with results from other groups). This 600-point fitting window was chosen based on the assumption that the earlier drift behavior is less dependent on any effects from the varying read fields. An approximately linear relationship is observed for the drift coefficient (ν) versus $1/KT$, with slightly lower drift coefficients and slope compared to our previous results as shown in Fig. 5b. This difference appears to be significant, despite the large error bars associated with both sets of measurements, and it is interesting to note that the earlier measurements were done on larger length and larger width devices (W × L ≈ 122 –142 nm × 360 – 500 nm), which may point to lower drift at smaller dimensions. Negligible drift has been shown in phase-change nanowires [40] and superlattice structures [41].

## Conclusions

We presented detailed device-level electrical characterization of 20 nm thick melt-quenched $Ge_2Sb_2Te_5$ line cells of similar dimensions (W × L ≈ 66-100 nm × 352-388 nm) between 80 K and 300 K. Room-temperature I-V measurements show a low-to-high field transition at ~19 MV/m in agreement with previous reports. Resistance drift measurements show significant drift coefficients down to 80 K, with significant time dependency and device-to-device variations, pointing to complex dynamic physical processes underlying resistance drift, in addition to structural relaxation, such as charge trapping and de-trapping. A deeper understanding of resistance drift can be achieved using device structures that result in self-limited melt-quench amorphization with nearly constant amorphized volumes (likely smaller and more confined cells), in which decoupling of the expected mechanical, electronic, and thermal mechanisms involved can be attempted.

## Acknowledgments

This work was partially supported by the US National Science Foundation through awards NSF 1710468 and 1711626. The devices were fabricated at IBM Watson Research Center under a Joint Study Agreement.